\documentclass[
aps,
prc,
showpacs,
superscriptaddress,
nofootinbib,
floatfix]
{revtex4}
\usepackage{graphicx,
longtable,
slashed}
\begin{document}
\title{Multi-isotope degeneracy of neutrinoless double beta decay mechanisms\\  
            in the quasi-particle random phase approximation}
%
\author{        Amand~Faessler}
\affiliation{   Institute of Theoretical Physics,
				University of Tuebingen, 
               72076 Tuebingen, Germany}
\author{        G.L.~Fogli}
\affiliation{   Dipartimento Interateneo di Fisica ``Michelangelo Merlin,'' 
               Via Amendola 173, 70126 Bari, Italy}
\affiliation{   Istituto Nazionale di Fisica Nucleare, Sezione di Bari, 
               Via Orabona 4, 70126 Bari, Italy}
\author{        E.~Lisi}
\affiliation{   Istituto Nazionale di Fisica Nucleare, Sezione di Bari, 
               Via Orabona 4, 70126 Bari, Italy}
%
%
\author{        A.M.~Rotunno}
\affiliation{   Dipartimento Interateneo di Fisica ``Michelangelo Merlin,'' 
               Via Amendola 173, 70126 Bari, Italy}
\author{        F.~\v{S}imkovic}
\affiliation{	Department of Nuclear Physics and Biophysics,
				Comenius University, 
				Mlynsk\'a dolina F1, SK--842 15 Bratislava, Slovakia}
\affiliation{	Bogoliubov Laboratory of Theoretical Physics, JINR, 
				141980 Dubna, Moscow region, Russia}

\begin{abstract}
We calculate nuclear matrix elements (NME) of neutrinoless double beta
decay ($0\nu\beta\beta$) in four different candidate nuclei ($^{76}$Ge, $^{82}$Se, $^{100}$Mo, $^{130}$Te)
within the quasiparticle random phase approximation (QRPA) and its uncertainties.
We assume (up to) four coexisting mechanisms for $0\nu2\beta$ decay, mediated by 
light Majorana neutrino exchange ($\nu$), heavy Majorana neutrino exchange ($N$),  
\hbox{$R$-parity} breaking supersymmetry ($\slashed{R}$), and squark-neutrino $(\tilde q)$, 
interfering either constructively or destructively with each other. 
We find that, unfortunately, current NME uncertainties appear to
prevent a robust determination of the relative contribution of each mechanism to the
decay amplitude, even assuming accurate measurements of decay lifetimes.
The near-degeneracy of $0\nu\beta\beta$ mechanisms  
is analyzed with simple algebraic techniques, which do not involve 
assumptions about the statistical distribution of errors. 
We discuss implications of such degeneracy on prospective searches for absolute neutrino masses.
\end{abstract}
\medskip
\maketitle

\section{Introduction \label{SecI}}

The discovery of neutrino masses and mixings \cite{Na10} has given new impetus to the
search for lepton number violation (LNV) and, in particular, for the process of
neutrinoless double beta decay ($0\nu\beta\beta$) \cite{Bi10},
\begin{equation}
(Z,\, A)\to (Z+2,\, A) + 2e^- \ ,
\end{equation}
in a variety of experiments using different $(Z,\,A)$ nuclei \cite{Av08}.
The process can be mediated by light Majorana neutrinos, as well as by alternative
(and possibly co-existing)  mechanisms invoking new particles and interactions beyond the Standard Model \cite{Ro10}.

In the case of a single LNV mechanism (labelled by an index $j$), 
the $0\nu\beta\beta$ decay half-life $T_i$ in a given nucleus $i=(Z,\,A)$ reads
\begin{equation}
\label{Ti}
T_i^{-1} = G_i^j\, |M_i^j\,\eta_j|^2\ ,
\end{equation}
where $G_i^j$ is a calculable phase-space factor, $M_i^j$ is the $0\nu\beta\beta$ nuclear
matrix element (NME), and $\eta_j$ is a dimensionless LNV parameter, characteristic of the particle physics model. 
In the case of coexisting mechanisms with identical phase space ($G_i^j\equiv G_i$), the above expression
is generalized as \cite{Si10}
\begin{equation}
\label{Tiplus}
T_i^{-1} = G_i\, \Big|\sum_j M_i^j\,\eta_j\Big|^2\ . 
\end{equation}
In the above equation, the $\eta_j$ parameters may take either sign (and, allowing CP violation, 
even complex phases), leading 
to constructive or destructive interference in the decay amplitude. 

In the absence of independent constraints on the LNV parameters,
the discrimination of different mechanisms requires that either the $\eta_j$'s are factorized out, or that
they are determined through the data themselves. For instance, assuming one mechanisms at a time, one
may analyze $\eta_j$-independent ratios of half-life data $\{T_i\}$ from different nuclei (typically
with a ${}^{76}\mathrm{Ge}$ datum at denominator) \cite{Gr02,Pe04,De07,Ge07}, 
possibly including joint $(T_i,\,T_k)$ 
probability distributions \cite{Fo09}. Alternatively, one may assume (at least) as many $\left\{T_i\right\}$ data 
as the number of considered mechanisms, and then solve Eq.~(\ref{Tiplus}) in terms of the $\eta_j$'s \cite{Si10}.%

The studies in \cite{De07,Ge07,Fo09} suggest that, at least in principle, a few
(possibly coexisting \cite{Si10}) mechanisms may be discriminated with accurate half-life data, 
provided that the associated NME 
uncertainties in different nuclei are ``small enough.''  
Attempts to quantify this statement in prospective scenarios 
have been made through MonteCarlo simulations with educated error guesses \cite{Ge07} or through likelihood
analyses \cite{Fo09} including error covariances \cite{Fa09}. However, apart from the difficulty in estimating 
``$1\sigma$''  theoretical errors, the approaches in \cite{Ge07,Fo09} had to rely upon a somewhat inhomogeneous
comparison of NME's taken from different publications or nuclear models; conversely, the NME's in \cite{Si10} were
all calculated in the same model, but 
their variations were not fully included in the analysis, although it was
noted that they could be relevant.   
In this work we adopt an approach where: (1) all NME's and their variations are estimated
with one and the same method, the quasiparticle random phase approximation (QRPA \cite{Fa98}) and
(2) the  resulting discrimination of different mechanisms is analyzed with simple algebraic (rather than
statistical) tools. 

In particular, the effect of theoretical uncertainties is evaluated herein by varying several 
QRPA inputs, which require extensive NME calculations. As a consequence, our analysis is restricted to 
a relatively small set of candidate nuclei and mechanisms. We consider four among the 
experimentally promising \cite{Av08}
candidate nuclei, labelled by the
index $i$:
\begin{equation}
\label{i}
i={}^{76}\mathrm{Ge},\,{}^{82}\mathrm{Se},\, {}^{100}\mathrm{Mo},\, {}^{130}\mathrm{Te}\, ,
\end{equation}
and four possible $0\nu\beta\beta$ decay processes, labelled by the index $j$,
\begin{equation}
\label{j}
j=\nu,\,N,\,\slashed{R},\,\tilde q\,,
\end{equation}
corresponding to   
light Majorana neutrino exchange \cite{Taka}, heavy Majorana neutrino exchange \cite{Gor1,Gor2,Pant}, 
$R$-parity breaking supersymmetry
mechanism \cite{Mo86,Ve87,Hi95,Fa96}, and squark-neutrino mechanism \cite{Ko98,Hi98,Pa99}, respectively. We then study solutions
to Eq.~(\ref{Tiplus}) in terms of prospective $\{T_i\}$ data. Unfortunately, 
we find that the reconstructed LNV parameters $\eta_j$ of the
four mechanisms to $0\nu\beta\beta$ decays are destabilized once different, currently admissible QRPA variants
are considered.  The effective degeneracy of mechanisms may
have significant implications on the interpretation of upcoming $0\nu\beta\beta$ results,
and on their interplay with other observations sensitive to absolute neutrino masses.

Our work is structured as follows. In Section~II we describe the LNV mechanisms
considered for $0\nu\beta\beta$ decay. In Section~III we discuss relevant aspects of 
the nuclear structure calculations. In Section~IV we analyze the problem of discriminating 
the mechanisms within QRPA uncertainties.
In Section~V we summarize our results and discuss their implications for absolute neutrino mass
searches.

\section{LNV mechanisms}

In this section we describe relevant features of the four LNV mechanisms in Eq.~(\ref{j}). In each
$i$-th nucleus [Eq.~(\ref{i})], they share the same phase space $G_i$, estimated in \cite{Pant} as
\begin{equation}
G({}^{76}\mathrm{Ge},\,{}^{82}\mathrm{Se},\, {}^{100}\mathrm{Mo},\, {}^{130}\mathrm{Te})=
(7.93,\,35.2,\,57.3,\,55.4)\times 10^{-15}\,\mathrm{y}^{-1}\ .
\end{equation}
Conventionally, the above phase space factors embed the fourth power of the axial coupling constant at
the fixed value $g_A=1.25$, and the inverse square of the nuclear radius $R=r_0 A^{1/3}$, with fixed
$r_0=1.1$~fm. They are compensated by factors $(g_A/1.25)^2$ and $1/R$ in the definition of 
the matrix elements $M_i^j$ \cite{Asse}.

\subsection{Light Majorana neutrino exchange ($\nu$)}

In the usual case of exchange of three light Majorana neutrinos $\nu_h$ (with masses $m_h$,
Majorana phases $\phi_h$, and $\nu_e$ mixings $U_{eh}$), the LNV parameter takes the well-known form
\begin{equation}
\eta_\nu = \sum_{h=1}^3 |U_{eh}|^2 e^{i\phi_h} \frac{m_h}{m_e}\equiv \frac{m_{\beta\beta}}{m_e}\ ,
\end{equation}
where $m_e$ is the electron mass, and $m_{\beta\beta}$ is the ``effective Majorana mass.''
In each nucleus, the nuclear matrix element $M^\nu$ consists of Fermi (F), Gamow-teller (GT) and tensor (T) contributions,
\begin{equation}
\label{Mnu}
M^\nu = \left(\frac{g_A}{1.25}\right)^2
\left(M^\nu_\mathrm{GT}+M^\nu_\mathrm{T}-\displaystyle\frac{M^\nu_\mathrm{F}}{g^2_A}\right)\ ,
\end{equation}
with F, GT, and T operators defined as in \cite{Pant,Anat}.

\subsection{Heavy Majorana neutrino exchange ($N$)}

The neutrino mass spectrum may include, in principle, any number of heavy Majorana states $N_k$ with
masses $M_k$ above the typical $0\nu\beta\beta$ energy scale, i.e., greater than $O(10)$~GeV. 
We restrict our consideration to the case
  of right-handed Majorana fermions, which are singlets
  of the  $\mathrm{SU}_L(2)\times \mathrm{U}(1)$ group. Then, heavy neutrinos can 
mediate $0\nu\beta\beta$ decay with a LNV parameter $\eta_N$ given by
\begin{equation}
\eta_N = \sum_{k}^{\mathrm{heavy}} |U_{ek}|^2 e^{i\Phi_k} \frac{m_p}{M_k}\ ,
\end{equation}
where $m_p$ is the proton mass, while the $U_{ek}$ are the mixing matrix elements associated to left-handed weak interactions.
Also in this case, 
the nuclear matrix element $M^N$ is of the form
\begin{equation}
\label{MN}
M^N = \left(\frac{g_A}{1.25}\right)^2
\left(M^N_\mathrm{GT}+M^N_\mathrm{T}-\displaystyle\frac{M^N_\mathrm{F}}{g^2_A}\right)\ ,
\end{equation}
with F, GT, and T operators discussed at length in \cite{Pant}.

\subsection{R-parity breaking mechanism ($\slashed{R}$)}

In supersymmetric (SUSY) models one may define a multiplicative quantum number $R=(-1)^{2S+3B+L}$
(with S, B, and L being the spin, baryon, and lepton numbers), which takes the value $+1$ for ordinary
particles, and $-1$ for their superpartners. $R$-parity breaking allows (among others) 
superpotential terms of the form $W_{\slashed{R}}\ni \lambda'_{ijk}L_i Q_j D_k^c$, where $L$ and $Q$ are
lepton and quark doublet left-handed superfields, $D^c$ is a down quark singlet superfield, and
$\lambda'$ is a trilinear coupling. These terms can trigger  $0\nu\beta\beta$ decay 
with short-range exchange of heavy superpartners (see below in this subsection) and with 
long-range exchange of heavy squarks plus light neutrinos (see the next subsection).

Assuming a dominant, short-range exchange of a heavy gluino with mass $m_{\tilde g}$, 
the LNV parameter $\eta_{\slashed R}$ takes a simplified form, symmetric in the masses of the
$u$ and $d$ squarks:
\begin{equation}
\eta_\slashed{R} = \frac{\pi\alpha_s}{6}\frac{(\lambda'_{111})^2}{G^2_F}\frac{m_p}{m_{\tilde g}}
\frac{(m^2_{\tilde u_L}+m^2_{\tilde d_R})^2}{m^4_{\tilde u_L}\cdot m^4_{\tilde d_R}}\ ,
\end{equation}
where $G_F$ is the Fermi constant and $\alpha_s$ the strong coupling constant. At hadron level,
in the hypothesis of one-pion and two-pion exchange dominance, the matrix element contains
both tensor and Gamow-Teller contributions, 
\begin{equation}
M^{\slashed{R}}=\left(\frac{g_A}{1.25}\right)^2
\left[c^{1\pi}(M_\mathrm{T}^{1\pi}+M^{1\pi}_\mathrm{GT})+c^{2\pi}(M_\mathrm{T}^{2\pi}+M^{2\pi}_\mathrm{GT})\right]\ ,
\end{equation}
with known prefactors $c^{1\pi}$ and $c^{2\pi}$. See \cite{Fa96,Wo99} for further details.

\subsection{Squark-neutrino mechanism ($\tilde q$).}

$R$-parity breaking SUSY may trigger $0\nu\beta\beta$ decay via long-range exchange of  (not necessarily massive) 
neutrinos plus squarks \cite{Gu08}. The corresponding LNV parameter is sensitive
to down-squark ($\tilde d_{(k)}=\tilde d, \,\tilde s,\,\tilde b$) masses and left-right mixings:
\begin{equation}
\eta_{\tilde q} = \sum_k \frac{\lambda'_{11k}\lambda'_{1k1}}{2\sqrt{2}G_F}\sin2\theta^d_{(k)}
\left( \frac{1}{m^2_{\tilde d_{1(k)}}}-\frac{1}{m^2_{\tilde d_{2(k)}}}\right).
\end{equation}
At hadron level, in the hypothesis of pion-exchange dominance, 
the matrix element contains
both tensor and Gamow-Teller contributions, 
with a resulting matrix element of the form
\begin{equation}
M^{\tilde q}=\left(\frac{g_A}{1.25}\right)^2
(M_\mathrm{GT}^{\pi}-M^{\pi}_\mathrm{T})\ .
\end{equation}
See  \cite{Gu08} for further details.

\subsection{Remarks on $\nu$, $N$, $\slashed R$, $\tilde q$ mechanisms}

The above four mechanisms are different in several respects: 
light and heavy neutrino exchange NME's do not exactly scale as 
$g_A^2$ (because of Fermi contributions), unlike SUSY mechanisms, which also have large
tensor contributions; moreover, short-range mechanisms (such as  
heavy neutrino exchange) are particularly sensitive to the nucleon-nucleon potential.
Thus, one might expect to find different NME isotopic patterns 
in different mechanisms. However, we shall see in Sec.~IV that the 
resulting differences in NME patterns are comparable to current
QRPA variations,
thus preventing an effective discrimination of the mechanisms
based on multi-isotope $0\nu\beta\beta$ half-life data. 

The similarity of NME patterns in the different nuclei and mechanisms considered  herein
suggests that nucleon (rather than nuclear) physics dominates in all cases. 
For the $\nu$ mechanism, the dominance of short-range nucleon-nucleon contributions 
has been shown in different approaches \cite{Anat,She1,Rath}. It would be useful
to perform similar studies for nonstandard mechanisms as well.

\section{Nuclear structure calculations}

In what follows, the $0\nu\beta\beta$ nuclear matrix elements $M_i^j$ are evaluated
(for $i={}^{76}\mathrm{Ge},\,{}^{82}\mathrm{Se},\, {}^{100}\mathrm{Mo},\, {}^{130}\mathrm{Te}$
and  $j=\nu,\,N,\,\slashed{R},\,\tilde q$) under the so-called self-consistent renormalized QRPA
(SRQRPA) \cite{De97,Kr98,Sim08}, which takes into account the Pauli 
exclusion principle and conserves the mean particle number
in correlated ground states. The following variants are introduced to account for intrinsic QRPA
uncertainties.  

For each nucleus, two choices of single-particle basis are considered: intermediate and large.
The intermediate-size model space (m.s.) has 12 levels (oscillator shells $N=2$--4)
for $^{76}$Ge, $^{82}$Se, 16 levels (oscillator shells $N=2$--4 plus 
the $f+h$ orbits from $N=5$) for $^{100}$Mo  and 18 levels 
(oscillator shells $N=3$,~4 plus $f+h+p$ orbits from $N=5$) for $^{130}$Te.
The large-size single particle space contains 21 levels 
(oscillator shells $N=0$--5)  for  $^{76}$Ge, $^{82}$Se, $^{100}$Mo, 
and 23 levels ($N=1$--5 and $i$ orbits from $N=6$) for $^{130}$Te.   
In comparison with previous studies \cite{Asse} we omit 
the small-size model space, as it is insufficient to describe 
realistically the tensor matrix elements.

The single particle energies are obtained by using
a Coulomb-corrected Woods-Saxon potential. Two-body $G$-matrix
elements are derived from the Argonne and the 
Charge Dependent Bonn (CD-Bonn) 
one-boson exchange potential within the Brueckner theory. 
The schematic pairing interactions are 
adjusted to fit the empirical pairing gaps \cite{Ch93}. The
particle-particle and particle-hole channels of the $G$-matrix
interaction of the nuclear Hamiltonian $H$ are renormalized by
introducing the parameters $g_{pp}$ and $g_{ph}$, respectively.
The calculations are been carried out for $g_{ph} = 1.0$.

The particle-particle strength parameter $g_{pp}$ 
of the SRQRPA is fixed by the data on the two-neutrino double 
beta ($2\nu\beta\beta$) decays \cite{Asse}.
In the calculation of the $0\nu\beta\beta$-decay NME's, we consider 
the two-nucleon short-range correlations  
derived from the Argonne or CD-Bonn nucleon-nucleon ($NN$) potentials 
as residual interactions \cite{St09}.
The measured $2\nu\beta\beta$-decay 
half-lives $T_{2\nu}$, taken from \cite{Ba09}, fix the matrix element  
$|M^{2\nu}_\mathrm{GT}|$ via $T^{-1}_{2\nu} = G_{2\nu} g^4_A |m_e M_\mathrm{GT}^{2\nu}|^2$,
where $G_{2\nu}$ is the $g_A$-independent phase space factor for $2\nu\beta\beta$ decay. The
value of $g_{pp}$ is thus fixed for any given $g_A$, which we take equal to
either its bare value ($g_A=1.25$) or its quenched value ($g_A=1.00$) for all nuclei. 
For simplicity, we neglect the possibility of different $g_A$ quenching in different nuclei \cite{Over}.
Finally, for each $2\nu\beta\beta$ constraint to $g_{pp}$,
we take both the  $T_{2\nu}$ central value and its $\pm1\sigma$ deviations,
leading to a triplet of $g_{pp}$ estimates (best fit $\pm1\sigma$) in each case.

Summarizing, our analysis is based on the calculation of 384 NME's, corresponding to four nuclei
($i$), four mechanisms ($j$), two model space sizes (m.s.), two nucleon-nucleon potentials ($NN$), two axial-coupling values
($g_A$), and three particle-particle strength values  $(g_{pp})$,
\begin{equation}
384 (\mathrm{NME})= 4 (i)\times 4(j)\times 2 (\mathrm{m.s.}) \times 2 (NN) \times 2 (g_A) \times 3 (g_{pp})\ .
\end{equation}
However, we have verified a posteriori that $g_{pp}$ variations within $\pm1\sigma$ induce NME variations
much smaller than those induced by changes in model space, nucleon potential, or $g_A$. The analysis
discussed in the next Section turns out to be basically unaffected by
such extra $g_{pp}$ variations. In what follows,
we shall thus focus on the numerical results obtained at the central values for the $g_{pp}$'s. 
In particular, we restrict our analysis to eight $4\times 4$ matrices $M_i^j$
of NME's, corresponding to eight QRPA variants, associated to changes in model space size, $NN$ potential,
and $g_A$,
\begin{equation}
8 (M_i^j)= 2 (\mathrm{m.s.}) \times 2 (NN) \times 2 (g_A) \ .
\end{equation}
Table~\ref{Matrices} reports the corresponding  $M_i^j$ estimates with three significant digits. 

\begin{table}[h]
\caption{\label{Matrices}
Matrices $M_i^j$ of NME's for 
$i=({}^{76}\mathrm{Ge},\,{}^{82}\mathrm{Se},\,{}^{100}\mathrm{Mo},\,{}^{130}\mathrm{Te})$
and $j=(\nu,\,N,\,\slashed{R},\,\tilde q)$, for each of the 8 variants adopted within the QRPA.
The $2\times2\times2=8$ variants correspond to different choices for
the model space size (intermediate or large), for the  $NN$ potential (Argonne or CD-Bonn), and for the
$g_A$ value (1 or 1.25).} 
\begin{ruledtabular}
\begin{tabular}{cccccc}
QRPA case  &  $i\backslash j$  & $\nu$ & $N$ & $\slashed{R}$ & $\tilde q$ \\
\hline
Variant n.\ 1:					& $^{76}$Ge & 3.85 & 172 & 387 & 396 \\ 
$g_A=1$,						& $^{82}$Se & 3.59 & 165 & 375 & 379 \\
$NN=$~Argonne,					&$^{100}$Mo & 3.62 & 185 & 412 & 405 \\
m.s.~=~intermediate				&$^{130}$Te & 3.29 & 171 & 385 & 382 \\
\hline
Variant n.\ 2: 					& $^{76}$Ge & 4.40 & 196 & 461 & 476 \\ 
$g_A=1$,						& $^{82}$Se & 4.19 & 193 & 455 & 465 \\
$NN=$~Argonne,					&$^{100}$Mo & 3.91 & 192 & 450 & 449 \\
m.s.~=~large					&$^{130}$Te & 3.34 & 177 & 406 & 403 \\
\hline
Variant n.\ 3: 					& $^{76}$Ge & 4.15 & 269 & 340 & 408 \\ 
$g_A=1$,						& $^{82}$Se & 3.86 & 259 & 329 & 390 \\
$NN=$~CD-Bonn,					&$^{100}$Mo & 3.96 & 299 & 356 & 416 \\
m.s.~=~intermediate				&$^{130}$Te & 3.64 & 277 & 336 & 397 \\
\hline
Variant n.\ 4: 					& $^{76}$Ge & 4.69 & 317 & 393 & 483 \\ 
$g_A=1$,						& $^{82}$Se & 4.48 & 312 & 388 & 472 \\
$NN=$~CD-Bonn,					&$^{100}$Mo & 4.20 & 311 & 384 & 455 \\
m.s.~=~large					&$^{130}$Te & 3.74 & 294 & 350 & 416 \\
\hline
Variant n.\ 5: 					& $^{76}$Ge & 4.75 & 233 & 587 & 594 \\ 
$g_A=1.25$,						& $^{82}$Se & 4.54 & 226 & 574 & 578 \\
$NN=$~Argonne,					&$^{100}$Mo & 4.40 & 250 & 629 & 612 \\
m.s.~=~intermediate				&$^{130}$Te & 4.16 & 234 & 595 & 589 \\
\hline
Variant n.\ 6: 					& $^{76}$Ge & 5.44 & 265 & 700 & 718 \\ 
$g_A=1.25$,						& $^{82}$Se & 5.29 & 263 & 698 & 710 \\
$NN=$~Argonne,,					&$^{100}$Mo & 4.79 & 260 & 690 & 683 \\
m.s.~=~large					&$^{130}$Te & 4.18 & 240 & 626 & 620 \\
\hline
Variant n.\ 7: 					& $^{76}$Ge & 5.11 & 351 & 515 & 612 \\ 
$g_A=1.25$,						& $^{82}$Se & 4.88 & 340 & 504 & 595 \\
$NN=$~CD-Bonn,					&$^{100}$Mo & 4.81 & 388 & 544 & 628 \\
m.s.~=~intermediate				&$^{130}$Te & 4.62 & 364 & 519 & 611 \\
\hline
Variant n.\ 8: 					& $^{76}$Ge & 5.82 & 412 & 596 & 728 \\ 
$g_A=1.25$,						& $^{82}$Se & 5.66 & 408 & 594 & 720 \\
$NN=$~CD-Bonn,					&$^{100}$Mo & 5.15 & 404 & 589 & 691 \\
m.s.~=~large					&$^{130}$Te & 4.70 & 384 & 540 & 641 
\end{tabular}
\end{ruledtabular}
\end{table}
\vfill

\section{Discrimination of mechanisms: Expectations and results}

The system of equations in~(\ref{Tiplus}) involves, in general, four unknown complex parameters $\eta_j$. For the sake
of simplicity, as in \cite{Si10}, we restrict ourselves to a CP-conserving scenario with real $\eta_j$. 
The relative sign of the $\eta_j$'s can, however, be positive or negative, leading to
constructive or destructive interference among different mechanisms. Taking square roots in
Eq.~(\ref{Tiplus}) leads to a further $\pm$ sign ambiguity for each index $i$ \cite{Si10}.
We restrict our analysis to the case with all ``$+$'' signs for each $i$, namely, to the linear system
of equations
\begin{equation}
\label{Linear}
\sum_{j=1}^4 M_i^j \eta_j = \mu_i\ (i=1,\dots,4)\ ,
\end{equation}
where 
\begin{equation}
\mu_i = (G_i T_i)^{-1/2}\ .
\end{equation}
Relaxing the above simplifying assumptions would corroborate the results which follow.

\subsection{Expectations}

For given multi-isotope  data $\{T_i\}$, solutions to the linear system in Eq.~(\ref{Linear}) 
in terms of $\{\eta_j\}$ should identify, in principle, the underlying physics
mechanism(s) \cite{Si10}. 
Testing different mechanisms, as well as their possible coexistence, becomes then
a parameter estimation problem in $\eta_j$-space, provided that 
theoretical uncertainties (for $M_i^j$) and experimental errors (for $T_i$) are accounted for. 
For instance, if $0\nu\beta\beta$ decays were due only to light Majorana neutrino
exchange, accurate solutions to Eq.~(\ref{Linear}) should ideally cluster around a point
 $(\eta_\nu,\,\eta_N,\,\eta_{\slashed{R}},\,\eta_{\tilde q}) \simeq 
(\langle \eta_\nu\rangle,\,0,\,0,\,0)$. Conversely,
nonzero values for two or more $\eta_j$'s would  signal the coexistence
of two or more mechanisms in $0\nu\beta\beta$ decay. This approach has been discussed in \cite{Si10} where, 
however, the NME values were taken from a single QRPA calculation with fixed input parameters, 
whose possible variations were not considered.

\begin{figure}[h]
\vspace*{1.0cm}
\hspace*{0.cm}
\includegraphics[scale=.97]{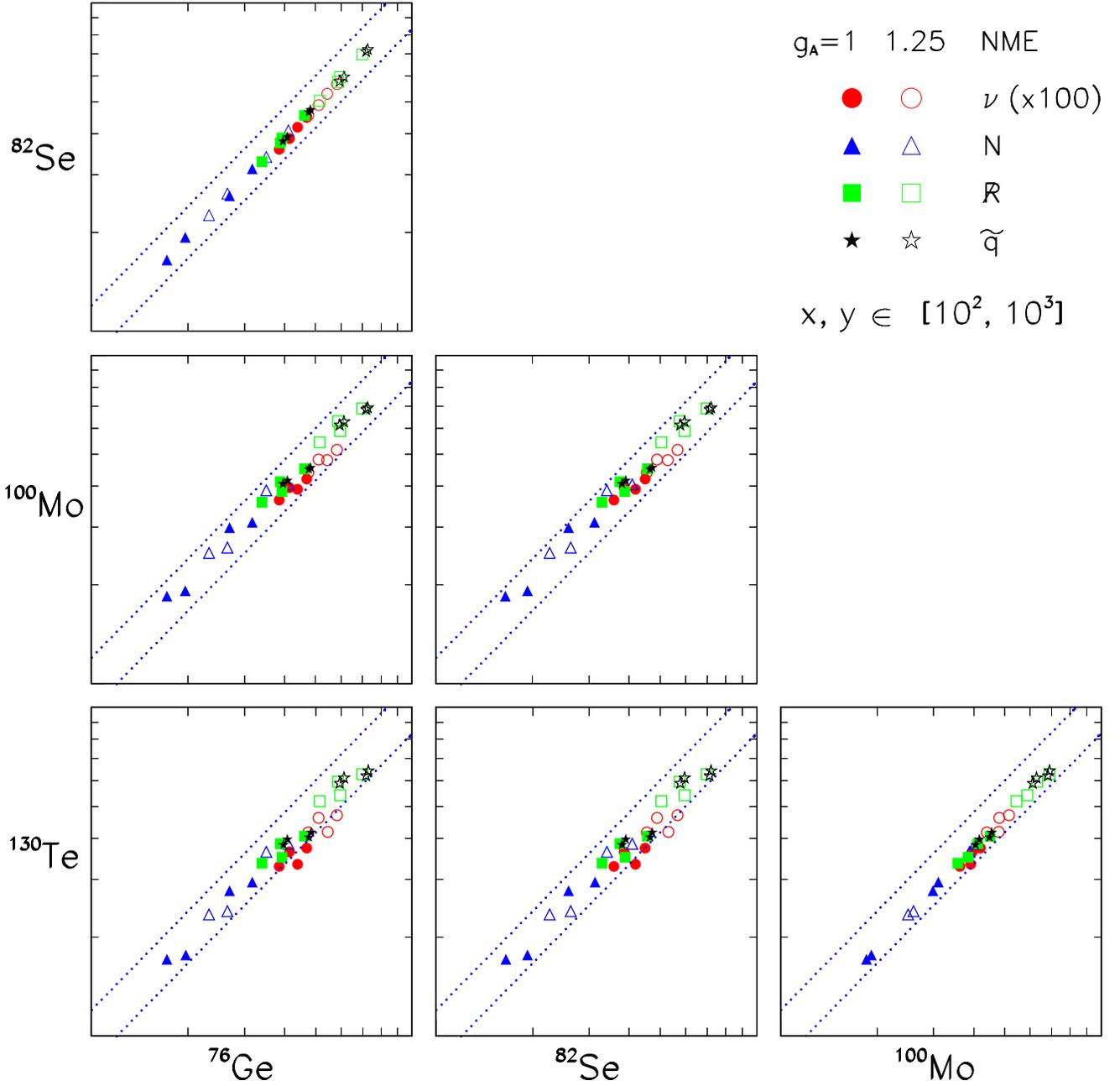}
\vspace*{+1.0cm}
\caption{ \label{f01} Graphical representation of the NME's in Tab.~\ref{Matrices} as a 
scatter plot, for each couple of nuclei 
among $^{76}$Ge, $^{82}$Se, $^{100}$Mo, and $^{130}$Te. Eight QRPA variants are
shown for each of the four different $0\nu\beta\beta$ mechanisms, 
labelled as $\nu$ (circles), $N$ (triangles), $\slashed{R}$ (squares), and $\tilde q$ (stars). 
Solid and empty markers refer to $g_A=1$ and $g_A=1.25$, respectively. 
In each panel, the $(x,y)$ coordinates span the range
$[10^2,\,10^3]$ in logarithmic scale. For the
$\nu$ mechanism, NME's are scaled ($\times 100$) to fit the range.    
Diagonal dotted lines mark the band where the $x$ and $y$ coordinates differ by up to $20\%$. 
The NME's appear to cluster along the diagonals, suggesting an effective degeneracy
of the $0\nu\beta\beta$ mechanisms. See the text for details.}
\end{figure}

As we show below, it turns out that the solutions 
$\{\eta_j\}$  are destabilized once QRPA variants are accounted for, even in the presence
of accurate measurements ($\delta T_i/T_i\ll 1$). In particular, 
the linear system of Eq.~(\ref{Linear}) is shown
to be ill-conditioned, with matrices $M_i^j$ of nearly rank one, leading to an effective
degeneracy of mechanisms.  
Before dealing with the $M_i^j$ algebra, it is 
useful to illustrate the source of such degeneracy in a graphical way. 

Figure~1 shows the contents of Tab.~\ref{Matrices} as a scatter plot of NME's,
for each couple of nuclei among $^{76}$Ge, $^{82}$Se, $^{100}$Mo, and $^{130}$Te. For each mechanism,
the eight QRPA variants correspond to eight markers  (circles
for $\nu$, triangles for $N$, squares for $\slashed{R}$, and stars for $\tilde q$), 
solid and empty markers corresponding to $g_A=1$ and $g_A=1.25$, respectively.
NME's for the $\nu$ mechanism are multiplied by a factor of 100 in order to fit in the
chosen logarithmic scales, which span the range $[10^2,\,10^3]$. Note that, on such scales, the effects of $g_{pp}$
variations mentioned in Sec.~III would be unobservable.
Diagonal dotted lines mark the band where the $x$ and $y$ coordinates are equal up to $20\%$. 
The source of degeneracy is then rather clear: all NME's cluster along the diagonal in each
panel, with little spread (less than about $20\%$) in the orthogonal direction. In particular,
the transverse spread induced by different mechanisms appears to be comparable to that induced by the
QRPA variants.
In other words, the columns of the $M_i^j$ matrices are roughly proportional
to each other, and with comparable entries for different nuclei, 
within QRPA uncertainties. The linear system in Eq.~(\ref{Linear}) is then expected
to be significantly degenerate, with nearly ``null determinant'' not only for the full $4\times4$ 
matrix $M_i^j$, but also for all of its $3\times 3$ and $2\times 2$ submatrices. 
These expectations are confirmed by the following numerical arguments.

\medskip
\subsection{Numerical results}
\medskip

Stability and degeneracy issues in linear equation systems are best dealt with the
Singular Value Decomposition (SVD) technique \cite{Alg1,Alg2}. For a square matrix $\mathbf{M}$ as in
Eq.~(\ref{Linear}) the SVD is equivalent to diagonalizing the product matrix $\mathbf{M}\cdot \mathbf{M}^T$,
taking the square roots of the eigenvalues, and sorting them in descending order. The resulting
set of ``singular values'' $\sigma_i$ encodes relevant information on the  stability of solutions \cite{Alg1,Alg2}. 
In fact, if all $\sigma_i$'s have comparable magnitudes, the $\eta_j$ solutions to 
Eq.~(\ref{Linear}) tend to be stable under small variations in either the $M_i^j$ or $\mu_i$ entries.
Conversely, if at least one $\sigma_i$ is much smaller than the others,
the system is ill-conditioned and generates unstable solutions. In particular, an ill-conditioned system, 
characterized by a large ``condition number'' $\kappa=\sigma_{\max}/\sigma_{\min}$, is expected to become 
unstable when the fractional variations $\delta M_i^j$ or $\delta \mu_i$ become comparable to the reciprocal condition
number  $\rho=1/\kappa$ \cite{Alg1,Alg2}. A slightly more technical point concerns the optimization of $\rho$.
Although $\rho$ is invariant under an overall
scaling of matrix entries ($M_i^j\to f\cdot M_i^j$), it may change under separate scalings of rows
and columns ($M_i^j\to r_i \cdot M_i^j \cdot s_j$), which transform Eq.~(\ref{Linear}) into an
equivalent system, provided that $\eta_j \to \eta_j/s_j$ and $\mu_i=\mu_i \cdot s_i$. The maximization of
$\rho$ by row/column rescalings, usually called ``matrix equilibration'' \cite{Alg1,Alg2,LAPA}, provides more
robust estimates of the fractional errors which may trigger unstable solutions.
Summarizing, the SVD of an equilibrated 
matrix provides useful information about the effective matrix rank
(equal to the number of ``large'' singular values) and about the size of ``dangerous'' fractional errors
(equal to the reciprocal condition number $\rho$).

Table~\ref{SVD} reports the singular values and 
(reciprocal) condition numbers, as obtained by applying  equilibration and SVD routines \cite{LAPA} to
the eight matrices $M_i^j$ in Tab.~\ref{Matrices}. In each of the eight QRPA variants, 
only one singular value is of $O(1)$, the others being smaller by one to three orders of magnitude.
The matrices $M_i^j$ are thus nearly of rank one, leading to basically undetermined and degenerate
solutions. Instabilities of solutions to Eq.~(\ref{Linear}) are expected 
to arise already for $\delta M_i^j$ as small as $\rho\sim O(10^{-3})$, implying that even
three-digit approximations are
potentially dangerous. 

\newpage
\begin{table}[t]
\caption{\label{SVD} Singular value decomposition of the (equilibrated) matrices $M_i^j$ given in Tab.~\protect\ref{Matrices}. 
For each of the eight QRPA variants, we report the 
four singular values $\sigma_i$ (in descending order), 
the matrix condition number $\kappa$, and the reciprocal condition number $\rho=1/\kappa$. 
The inequality $\sigma_1\gg \sigma_{2,3,4} $ signals that each matrix $M_i^j$ is ill-conditioned and
nearly threefold degenerate.
See the text for details.}
\begin{ruledtabular}
\begin{tabular}{ccccccc}
Variant & $\sigma_1=\sigma_{\max}$ & $\sigma_2$ & $\sigma_3$ & $\sigma_4=\sigma_{\min}$ & 
$\kappa=\sigma_{\max}/\sigma_{\min}$ & $\rho = \sigma_{\min}/\sigma_{\max}$ \\
\hline
1 & 3.91 & $9.92\times 10^{-2}$ & $14.3\times 10^{-3}$ & $4.71\times 10^{-3}$ & $0.83\times 10^3$ & $1.20\times 10^{-3}$ \\
2 & 3.91 & $9.66\times 10^{-2}$ & $7.46\times 10^{-3}$ & $3.40\times 10^{-3}$ & $1.15\times 10^3$ & $0.87\times 10^{-3}$ \\
3 & 3.89 & $9.66\times 10^{-2}$ & $21.4\times 10^{-3}$ & $3.50\times 10^{-3}$ & $1.11\times 10^3$ & $0.90\times 10^{-3}$ \\
4 & 3.91 & $8.26\times 10^{-2}$ & $13.6\times 10^{-3}$ & $3.58\times 10^{-3}$ & $1.09\times 10^3$ & $0.92\times 10^{-3}$ \\
5 & 3.91 & $10.6\times 10^{-2}$ & $14.3\times 10^{-3}$ & $2.86\times 10^{-3}$ & $1.37\times 10^3$ & $0.73\times 10^{-3}$ \\
6 & 3.91 & $10.0\times 10^{-2}$ & $8.20\times 10^{-3}$ & $1.28\times 10^{-3}$ & $3.05\times 10^3$ & $0.33\times 10^{-3}$ \\
7 & 3.89 & $10.0\times 10^{-2}$ & $17.3\times 10^{-3}$ & $5.74\times 10^{-3}$ & $0.68\times 10^3$ & $1.48\times 10^{-3}$ \\
8 & 3.91 & $8.68\times 10^{-2}$ & $16.9\times 10^{-3}$ & $2.04\times 10^{-3}$ & $1.92\times 10^3$ & $0.52\times 10^{-3}$ \\
\end{tabular}
\end{ruledtabular}
\end{table}

Let us discuss a representative 
numerical check, by assuming a scenario with true NME's taken from the first matrix of Tab.~\ref{Matrices},
and true LNV parameters $(\eta_\nu,\,\eta_N,\,\eta_{\slashed{R}},\,\eta_{\tilde q})$ set to  
$(6.18\times 10^{-7},\,0,\,0,\,0)$, corresponding to a single $0\nu\beta\beta$ mechanism, namely, 
light Majorana neutrino exchange with $m_{\beta\beta}=0.316$~eV.
Such choices for $M_i^j$ and $\eta_j$ imply the following $\mu_i$ values (up to three significant digits):
\begin{equation}
\label{mudata}
\mu ({}^{76}\mathrm{Ge},\,{}^{82}\mathrm{Se},\, {}^{100}\mathrm{Mo},\, {}^{130}\mathrm{Te})=
(2.38,\,2.22,\,2.24,\,2.03)\times 10^{-6}\ ,
\end{equation}
and a $^{76}$Ge half-life of $2.23\times 10^{25}$~y, which  purposely reproduces the 
claimed measurement in \cite{Kl04,Kl06}. With the above $\mu_i$'s, we solve Eq.~(\ref{Linear}) in the unknowns
$\eta_{j}$ for each of the eight matrices in Tab.~\ref{Matrices}, including the first one
(where we expect to recover the ``true'' $\eta_j$ by construction).

Table~\ref{Degenerate} shows the results of our exercise, in terms of the relative contribution $W_j$ of 
each $j$-th mechanism to the $0\nu\beta\beta$ amplitude in $^{76}$Ge ($i=1$),
\begin{equation}
W_j = \frac{M_1^j\eta_j}{\displaystyle\sum_{j=1}^4 M_1^j\eta_j}\ \Rightarrow\ \sum_{j=1}^4 W_j=1\ .
\end{equation}

\begin{table}[b]
\caption{\label{Degenerate}
Relative contribution $W_j$ of the $j$-th $0\nu\beta\beta$ mechanism 
to the total $^{76}$Ge  decay amplitude ($\sum_j W_j=1$), for eight QRPA variants, assuming 
$W_j=\delta_{j1}$ as ``true'' case, and 
precisely known  lifetimes $T_i$ for the four nuclei  (see the text for details). }
\begin{ruledtabular}
\begin{tabular}{crrrr}
Variant & $W_\nu$ & $W_N$ & $W_{\slashed R}$ & $W_{\tilde q}$ \\
\hline
1 & $+1.05$ & $-0.01$ & $+0.12$ & $-0.16$ \\
2 & $+3.14$ & $+5.62$ & $-1.05$ & $-6.71$ \\
3 & $+1.26$ & $-0.24$ & $+0.94$ & $-0.96$ \\
4 & $+3.83$ & $+2.63$ & $+2.13$ & $-7.59$ \\
5 & $+1.00$ & $+3.07$ & $-2.83$ & $-0.24$ \\
6 & $+7.90$ & $+16.22$ & $+1.86$ & $-24.98$ \\
7 & $+1.97$ & $+1.09$ & $-0.27$ & $-1.79$ \\
8 & $+6.64$ & $+5.90$ & $+5.99$ & $-17.53$ \\
\end{tabular}
\end{ruledtabular}
\end{table}

In the first row of Tab.~\ref{Degenerate}, one would expect to recover the true $\nu$ mechanism, corresponding
to $W_\nu=1$ and $W_{N,\slashed{R},\tilde q}=0$. However, this is not the case:
the three-digit approximation is already sufficient to generate  spurious contributions
(e.g., $W_{\tilde q}=-0.16$).  
In the second to eighth row in 
Tab.~\ref{Degenerate}, where QRPA variants entail relative variations $\delta M_i^j \gg O(10^{-3})$, 
disparate solutions emerge, with constructive or destructive interference of (sub)dominant mechanisms. 
For instance, in the eight row, three large coherent contributions from the $\nu$, $N$, and $\slashed R$ mechanism
are almost cancelled by an even larger opposite contribution by the $\tilde q$ mechanism. 
In Tab.~\ref{Degenerate}, the largest amplitude corresponds to $W_\nu$ in only two cases (1st and 3rd variant),
while in others it corresponds to either $W_N$ (5th variant) or $W_{\tilde q}$ (remaining variants). 

Note that the instability of the
$W_j$'s (in both magnitude and sign) arises from intrinsic properties of the NME matrices $M_i^j$ 
and would be present
also for other choices of the ``true'' LNV parameters (e.g. by assuming another
mechanism as dominant). Moreover, 
varying 
input data in Eq.~(\ref{mudata}) within prospective errors,
as well as removing the simplifying assumptions preceding Eq.~(\ref{Linear}),  
would only generate further instabilities or multiplicities in the solutions.

If the problem is reduced to just three or two mechanisms (instead of four),
the instabilities, although less severe, remain harmful. In 
generic $3\times 3$ and $2\times 2$ minors of the $M_i^j$ matrices, unstable solutions are 
expected to arise for fractional variations $\delta M_i^j$ of order
$\rho_3=\sigma_3/\sigma_1\sim \mathrm{few}\times 10^{-3}$ and 
$\rho_2=\sigma_2/\sigma_1\sim \mathrm{few}\times 10^{-2}$, respectively. 
Since current QRPA variants induce 
$\delta M_i^j\sim O(10^{-1})$, a robust reconstruction of the
``true'' scenario cannot be achieved.

As a numerical check, we have repeated the previous exercise 
in the subcase $({}^{76}\mathrm{Ge},\,{}^{130}\mathrm{Te})\times (\nu,\,\tilde q)$, which
is among the ``less degenerate'' $2\times 2$ ones.  
We find reciprocal condition numbers $\rho_{2\times2}\simeq \mathrm{few}~\%$, as expected.
Since QRPA variants entail $\delta M_i^j\sim O(10^{-1})$, solutions are unstable, and
the reconstruction of true case 
($W_\nu=1$) yields fake values for $W_\nu$, ranging from a minimum of 
0.23 (corresponding to dominant $\tilde q$ mechanism) to a maximum of 1.5 (corresponding to destructive
interference of $\nu$ and $\tilde q$ mechanisms). If we further change either 
$\mu({}^{76}\mathrm{Ge})$ or $\mu({}^{130}\mathrm{Te})$ by  $\pm 10\%$ (corresponding to $\pm 20\%$
experimental uncertainties in $0\nu\beta\beta$ half-lives), 
we find reconstructed values of $W_\nu$  
within a wider interval, $W_\nu\in [-1.4,\,+2.9]$. This range spans disparate solutions,
including coexisting $\nu$ and $\tilde q$ mechanisms with constructive or destructive interference,
as well as subcases close to single
$\nu$ or single $\tilde q$ mechanism ($W_\nu \simeq 1$ or $W_{\tilde q}\simeq 1$, respectively).
Analogous results hold
for other $2\times 2$ minors. Note that, since the
$M_i^j$ matrices are nearly of rank one $(\sigma_1\gg \sigma_{2,3,4})$, the degeneracy 
applies to rectangular sub-matrices as well; thus,
the situation would not significantly 
improved  by over-constraining the system, e.g., by forcing two mechanisms to (approximately)
fit more than two lifetime data among those considered. 

We stress that the observed degeneracy derives from the nearly threefold singularity of 
the matrices $M_i^j$, and not from 
specific choices of prospective input data $\mu_i$ or of true LNV parameters $\eta_{j}$. 
The phenomenological indistinguishability of the 
$(\nu,\,N,\,\slashed{R},\,\tilde q)$ mechanisms in the ($^{76}$Ge, $^{82}$Se, $^{100}$Mo, $^{130}$Te) nuclei
appears to be an unfortunate but general result, as far as the current QRPA framework and its estimated
variants [leading to $\delta M_i^j\sim O(10^{-1})$] are assumed.  The numerical analysis  
suggests a very ambitious accuracy goal  for prospective QRPA
estimates [i.e., a relative spread $\delta M_i^j$ at the few percent level], 
in order to clearly discriminate the relative contributions of 
two mechanisms by means of two nuclei, among those considered herein.

\medskip
\subsection{Discussion}
\medskip

The difficulty of separating the $\nu$ mechanism from heavy Majorana neutrino exchange or from SUSY-mediated decays
in multi-isotope $0\nu\beta\beta$ data  
had been noted earlier in \cite{De07,Ge07,Fo09}. The
results presented in this work corroborate such findings, with the advantage of being
based on a homogeneous comparison of NME's, calculated in 
well-defined variants of one and the same nuclear model,  for all the nuclei 
and mechanisms considered---which was not the case in \cite{De07,Ge07,Fo09}. Therefore, the 
emerging degeneracy appears to be quite general and not accidental, in the context of the considered mechanisms:
for any of them, the NME's turn out to have a similar size in the various nuclei, with uncertainties 
mostly absorbed by common scaling factors. 
The resulting numerical degeneracy in the determination of LFV parameters
is supported and quantified by simple algebraic arguments, which 
do not assume specific distributions of theoretical errors. As such, the analysis method
could be extended to other nuclei, mechanisms or nuclear structure models.

Taken together, the results of \cite{De07,Ge07,Fo09} and of this
paper show, from different viewpoints,
that  $0\nu\beta\beta$ mechanisms mediated by (light or heavy) neutrino exchange and 
by $\slashed{R}$ supersymmetry  may be phenomenologically degenerate in some nuclei of interest,
within realistic uncertainties. 
Therefore, $0\nu\beta\beta$ signals or limits
from such nuclei could be generically interpreted in terms of unconstrained linear combinations of 
coexisting ($\nu,\,N,\,\slashed{R},\,\tilde q$) mechanisms, including subcases with one dominant 
(if not single) mechanism. Breaking the degeneracy would require significantly more accurate
NME estimates than can be obtained in current, state-of-the-art QRPA. 

 It should also be noted that we have considered only CP-conserving cases [i.e., real $\mu_i$ in Eq.~(\ref{Linear})]
and, furthermore, we have discarded the discrete ambiguities related to the two possible signs 
 for each index $i$ ($\mu_i \to \pm \mu_i$)  \cite{Si10}, which would worsen the degeneracy problem by increasing
 the multiplicity of solutions. 
In CP-violating scenarios, with unknown relative phases $\phi_i$ among
different $0\nu\beta\beta$ decay amplitudes, 
the discrete ambiguities would become continuous ($\mu_i \to e^{i\phi_i} \mu_i$),
thus rendering the system in Eq.~(\ref{Linear}) undetermined, unless extra independent data were added
to constrain the unknowns $\phi_i$. In any case, removal of our simplifying 
assumptions about the CP properties of the $\mu_i$ parameters would make the degeneracy problem more (not less)
difficult to solve.

These  negative findings do not exclude, of course, more favorable scenarios in other nuclei or 
mechanisms. For instance, 
it has been noted in \cite{De07,Ge07,Fo09} that left-right symmetric models with hadronic right-handed currents
have better chances to be distinguished from $\nu$ exchange via multi-isotope
half-life data. Extra-dimensional models with Kaluza-Klein neutrino states may also
provide phenomenologically distinctive patterns  \cite{De07}. Moreover, 
additional information  from
angular and energy electron distributions in the final state
might help breaking degeneracies (see, e.g., \cite{Bori}).  
The analysis of further (potentially more promising) mechanisms and observables, and of their
relative discrimination in the presence of  theoretical variants, 
is left to future works.

 A final remark is in order. Besides the 
QRPA framework adopted in this work, other independent approaches have been
(and are being) adopted in the calculation of NME's for different nuclei, especially in the reference
case of light Majorana neutrino exchange. Recent notable contributions include, e.g., applications
of the Large Scale Shell Model (LSSM) \cite{LSSM}, of the Interacting Boson Model (IBM) 
\cite{Iach}, of the Projected Hartree-Fock-Bogoliubov method (PHFB) \cite{HFB1}, and of the Energy Density Functional 
method (EDF) \cite{HFB2}; see the review
in \cite{Fa11}. The comparison of results obtained with these and other methods (or their variants) 
shows that the current
spread of independent, state-of-the-art NME calculations exceeds the 
uncertainties estimated within the QRPA approach alone \cite{Fa11}, suggesting that further efforts are needed to constrain 
various theoretical approximations and systematics. These efforts, which would be boosted by
future experimental observations of $0\nu\beta\beta$ decay,
 will certainly help to understand better the
source(s) of the degeneracy discussed in this work and, hopefully, to reduce the NME 
uncertainties at values small enough to allow the discrimination of at least two mechanisms.

\section{Summary and implications}

We have considered nuclear matrix elements for neutrinoless double beta decay in four nuclei
($^{76}$Ge, $^{82}$Se, $^{100}$Mo, $^{130}$Te), within the self-consistent renormalized QRPA
approach, for four $0\nu2\beta$ decay mechanisms, mediated by 
light Majorana neutrino exchange ($\nu$), heavy Majorana neutrino exchange ($N$),  
$R$-parity breaking supersymmetry ($\slashed{R}$), and squark-neutrino $(\tilde q)$.
QRPA uncertainties have been evaluated by changing the model space size, the nucleon-nucleon
potential, and the axial coupling. We have found that, within current QRPA uncertainties, 
the four mechanisms appear to be largely degenerate, and could co-exist in phenomenologically
indistiguishable linear combinations (regardless of experimental errors and of possible complex phases). 
The degeneracy analysis is based on simple algebraic (rather than statistical) tools,
which could be extended to NME estimates for other nuclei or mechanisms.

As far as the degeneracy in the ($^{76}$Ge, $^{82}$Se, $^{100}$Mo, $^{130}$Te) nuclei is 
concerned, one could, e.g., interpret their $0\nu\beta\beta$ decay rates 
in terms of an effective phenomenological parameter of the form
\begin{equation}
\label{last}
m'_{\beta\beta} \simeq  m_{\beta\beta} + \Delta m_{\beta\beta} ,
\end{equation}
where $m_{\beta\beta}$ is the usual effective Majorana mass for light $\nu$ exchange, 
while $\Delta m_{\beta\beta}$ embeds possible extra
contributions from the $(N,\,\slashed{R},\,\tilde q)$ mechanisms, with either constructive $(\Delta m_{\beta\beta}>0)$
or destructive $(\Delta m_{\beta\beta}<0)$ interference with the $\nu$ mechanism. 
Note that, in the hypothetical case of nearly exact destructive 
interference ($m'_{\beta\beta}\simeq 0$), the process of $0\nu\beta\beta$ decay might be strongly
suppressed and even escape 
observation in the four nuclei, despite the presence of Majorana neutrinos with $m_{\beta\beta} > 0$.

A nonstandard degree of freedom $\Delta m_{\beta\beta}$ could be relevant in comparing $0\nu\beta\beta$ results
with other searches for absolute $\nu$ masses, such as $\beta$-decay (probing
$m^2_\beta={ \sum_h} |U_{eh}|^2 m^2_h$) and precision cosmology (probing $\Sigma = m_1+m_2+m_3$), especially
if conflicting data emerge. An example is suggested by current results.
Roughly speaking, for nearly degenerate neutrino masses 
($m_{1,2,3}\simeq m_\nu$) it is
$\Sigma \simeq 3m_\nu$, with typical cosmological constraints ($\Sigma\lesssim0.6$~eV \cite{Fo08}) placing the
upper bound $m_\nu \lesssim 0.2$~eV. Assuming only the standard $\nu$ mechanism, one has 
$m_{\beta\beta}\simeq m_\nu f$ ($f\in [0.38,\,1]$) which, if $m_{\beta\beta}\sim 0.3$~eV is inferred 
from the claim in \cite{Kl04,Kl06}, translates into a lower bound
$m_\nu \simeq m_{\beta\beta}/f \gtrsim 0.3$~eV.  
This well-known tension \cite{Fo08} could be relaxed
by interpreting the $0\nu\beta\beta$ claim, e.g., in terms of coexisting standard and nonstandard
mechanisms with  $m_{\beta\beta}\lesssim  0.2$~eV and $\Delta m_{\beta\beta}\gtrsim\!0.1$~eV, respectively
(but adding up to $m'_{\beta\beta}\sim 0.3$~eV); the same adjustment should then apply,
within QRPA uncertainties, to 
the associated $0\nu\beta\beta$ signals in 
$^{82}$Se, $^{100}$Mo, and $^{130}$Te.

We emphasize that our results refer to a specific set of four nuclei and four mechanisms, 
described within state-of-the-art QRPA nuclear structure calculations. 
Non-degenerate scenarios may well emerge for other $0\nu\beta\beta$ nuclei, mechanisms and  observables, 
or if prospective  NME estimates in QRPA (or in alternative nuclear models) 
can reach significantly higher accuracies. 
We plan to explore some of these possibilities in future works.

\newpage
\acknowledgments

\vspace*{-2mm}

The work of G.L.F, E.L., and A.M.R.\ is supported by the Italian Istituto Nazionale di Fisica 
Nucleare (INFN) and Ministero dell'Istruzione, dell'Universit\`a e della Ricerca 
(MIUR) through the ``Astroparticle Physics'' 
research project. A.F.\ and F.\v{S}.\ acknowledge support of the 
Deutsche Forschungsgemeinschaft within the project
436 SLK 17/298.  The work of F.\v{S}.\ was
also partially supported by the VEGA Grant agency of
the Slovak Republic under the contract N.~1/0639/09.

\textbf{\textit{Note Added}}. The NME estimates used in this work for various mechanisms and QRPA variants 
                  (see Table~\ref{Matrices}) are also being
                  reported in another recent preprint \cite{Pe11}, where they are analyzed 
                  within different scenarios for coexisting mechanisms 
                  (with detailed discussions of non-interfering cases \cite{Supp} and/or of complex $\eta$'s), 
                  as well as with different aims and perspectives. The approaches used
                  in our paper and in \cite{Pe11} are largely
                  non-overlapping, apart from the description of common NME inputs.
                  The paper \cite{Pe11} analyzes what one can learn about the importance of the 
                  different mechanisms for the neutrinoless double beta decay, in particular, if accurate enough data and 
                  matrix elements will be available. The present paper shows that the current accuracy of 
                  the matrix elements does not prevent degeneracies and needs to be improved. 
                  One will also need future data with smaller errors for 
                  $2\nu\beta\beta$ (to better constrain $g_{pp}$) and precise enough half lives 
                  for $0\nu\beta\beta$ decays to discriminate between different mechanisms for the neutrinoless decay.                   
                  Finally, concerning non-interfering cases \cite{Supp}, it should be noted that the associated
                  equations would be of the form $\Sigma_i (M^j_i)^2 \eta_j^2=\mu_i^2$
                  in the unknown $\eta^2_j$ \cite{Pe11}. Numerically,
                  we find that the $(M^j_i)^2$ matrices and sub-matrices  have reciprocal condition numbers
                  about twice as large as the $M^j_i$ (sub)matrices. However, since 
                  fractional differences due to different QRPA variants also double when $M_i^j\to (M_i^j)^2$, 
                  there is no net gain, and the $\eta_j$ solutions remain unstable within current uncertainties.

\vspace*{-2mm}



\begin{thebibliography}{99}


\bibitem{Na10} 	K.~Nakamura and S.T.~Petcov, ``Neutrino mass, mixing, and oscillations,'' in
					K.~Nakamura {\em et al.} (Particle Data Group), J.\ Phys.\ G {\bf 37}, 075021 (2010). 


\bibitem{Bi10}	S.~M.~Bilenky,
  				``Neutrinoless double beta-decay,''
  				Phys.\ Part.\ Nucl.\  {\bf 41}, 690 (2010)
  				[arXiv:1001.1946 [hep-ph]].


\bibitem{Av08}  F.~T.~Avignone~III, S.~R.~Elliott, and J.~Engel,
    			``Double Beta Decay, Majorana Neutrinos, and Neutrino Mass,''
               	Rev.\ Mod.\ Phys.\  {\bf 80}, 481 (2008)
               	[arXiv:0708.1033 [nucl-ex]].

\bibitem{Ro10} 	W.~Rodejohann,
					``Neutrinoless Double Beta Decay in Particle Physics,''
					in the Proceedings of {em Neutrino 2010}, 
					XXIV International Conference on Neutrino Physics and Astrophysics
					(Athens, Greece, 2010), Nucl.\ Phys.\ {\bf B} (Proc.\ Suppl.), to appear
					[arXiv:1011.4942 [hep-ph]].

\bibitem{Si10}	F.~\v{S}imkovic, J.~Vergados, and A.~Faessler,
  				``Few active mechanisms of the neutrinoless double beta-decay and effective mass of Majorana neutrinos,''
  				Phys.\ Rev.\  {\bf D82}, 113015 (2010).
				

\bibitem{Gr02}  S.~M.~Bilenky and J.~A.~Grifols,
 				``The possible test of the calculations of nuclear matrix elements of the
 				$0\nu\beta\beta$ decay,''
 				Phys.\ Lett.\  B {\bf 550}, 154 (2002)
 				[arXiv:hep-ph/0211101].

\bibitem{Pe04}  S.~M.~Bilenky and S.~T.~Petcov,
 				``Nuclear matrix elements of $0\nu\beta\beta$-decay: Possible test of the calculations,''
 				arXiv:hep-ph/0405237.

\bibitem{De07}	F.~Deppisch and H.~P\"as,
 				``Pinning down the mechanism of neutrinoless double beta decay with
 				measurements in different nuclei,''
 				Phys.\ Rev.\ Lett.\  {\bf 98}, 232501 (2007).

\bibitem{Ge07}	V.~M.~Gehman and S.~R.~Elliott,
 				``Multiple-isotope comparison for determining $0\nu\beta\beta$ decay
 				mechanisms,''
 				J.\ Phys.\ G {\bf 34}, 667 (2007)
 				[Erratum-ibid.\  {\bf G35}, 029701 (2008)]
 				[arXiv:hep-ph/0701099].

\bibitem{Fo09}	G.~L.~Fogli, E.~Lisi, and A.~M.~Rotunno,
  				``Probing particle and nuclear physics models of neutrinoless double beta decay with different nuclei,''
  			    Phys.\ Rev.\  {\bf D80}, 015024 (2009)
  				[arXiv:0905.1832 [hep-ph]].



\bibitem{Fa09}  A.~Faessler, G.~L.~Fogli, E.~Lisi, V.~Rodin, A.~M.~Rotunno, and F.~\v{S}imkovic,
  				``QRPA uncertainties and their correlations in the analysis of $0\nu\beta\beta$ decay,''
  				Phys.\ Rev.\  {\bf D79}, 053001 (2009).
  				[arXiv:0810.5733 [hep-ph]].


\bibitem{Fa98}	A.~Faessler and F.~\v{S}imkovic,
				``Double beta decay,''
 				J.\ Phys.\ G {\bf 24}, 2139 (1998)
                [arXiv:hep-ph/9901215].

\bibitem{Taka}	 M.~Doi, T.~Kotani, and E.~Takasugi,
 				``Double Beta Decay And Majorana Neutrino,''
 				Prog.\ Theor.\ Phys.\ Suppl.\  {\bf 83}, 1 (1985).


\bibitem{Gor1}	  R.~N.~Mohapatra and G.~Senjanovic,
                 ``Neutrino Masses And Mixings In Gauge Models With Spontaneous Parity Violation,''
                   Phys.\ Rev.\  D {\bf 23}, 165 (1981).

\bibitem{Gor2}     V.~Tello, M.~Nemevsek, F.~Nesti, G.~Senjanovic, and F.~Vissani,
  			``Left-Right Symmetry: from LHC to Neutrinoless Double Beta Decay,''
  				Phys.\ Rev.\ Lett.\  {\bf 106}, 151801 (2011)
  				[arXiv:1011.3522 [hep-ph]].

\bibitem{Pant}	 F.~\v{S}imkovic, G.~Pantis, J.~D.~Vergados, and A.~Faessler,
 				``Additional nucleon current contributions to neutrinoless double beta
 				decay,''
 				Phys.\ Rev.\  C {\bf 60}, 055502 (1999)
 				[arXiv:hep-ph/9905509].


\bibitem{Mo86}  R.~N.~Mohapatra,
  				``New Contributions to Neutrinoless Double beta Decay in Supersymmetric Theories,''
  				Phys.\ Rev.\  {\bf D34}, 3457-3461 (1986).

\bibitem{Ve87}	J.~D.~Vergados,
  				``Neutrinoless Double Beta Decay Without Majorana Neutrinos In Supersymmetric Theories,''
  				Phys.\ Lett.\  {\bf B184}, 55 (1987).

\bibitem{Hi95}	M.~Hirsch, H.~V.~Klapdor-Kleingrothaus, and S.~G.~Kovalenko,
  				``New constraints on R-parity broken supersymmetry from neutrinoless double beta decay,''
  				Phys.\ Rev.\ Lett.\  {\bf 75}, 17-20 (1995).


\bibitem{Fa96}  A.~Faessler, S.~Kovalenko, F.~\v{S}imkovic, and J.~Schwieger,
  				``Dominance of pion exchange in R-parity violating supersymmetry contributions to 
				 neutrinoless double beta decay,''
  				Phys.\ Rev.\ Lett.\  {\bf 78}, 183-186 (1997)
  				[arXiv:hep-ph/9612357].
				

\bibitem{Ko98}	A.~Faessler, S.~Kovalenko, and F.~\v{S}imkovic,
  				``Pions in nuclei and manifestations of supersymmetry in neutrinoless double beta decay,''
  				Phys.\ Rev.\  {\bf D58}, 115004 (1998)
  				[arXiv:hep-ph/9803253].				
				
\bibitem{Hi98}  M.~Hirsch and J.~W.~F.~Valle,
  				``Neutrinoless double beta decay in supersymmetry with bilinear R parity breaking,''
  				Nucl.\ Phys.\  {\bf B557}, 60-78 (1999).
  				[arXiv:hep-ph/9812463].				

\bibitem{Pa99} 	H.~P\"as, M.~Hirsch, H.~V.~Klapdor-Kleingrothaus,
  				``Improved bounds on SUSY accompanied neutrinoless double beta decay,''
  				Phys.\ Lett.\  {\bf B459}, 450-454 (1999)
  				[arXiv:hep-ph/9810382].			
				

\bibitem{Asse}   V.~A.~Rodin, A.~Faessler, F.~\v{S}imkovic and P.~Vogel,
  				``Uncertainty in the $0\nu\beta\beta$ decay nuclear matrix elements,''
  					Phys.\ Rev.\  {\bf C68}, 044302 (2003)
  				[arXiv:nucl-th/0305005];
                ``Assessment of uncertainties in QRPA $0\nu\beta\beta$-decay nuclear matrix
                elements,''
                Nucl.\ Phys.\  A {\bf 766}, 107 (2006)
                [Erratum-ibid.\  A {\bf 793}, 213 (2007)]
                [arXiv:0706.4304 [nucl-th]].

\bibitem{Anat}  F.~\v{S}imkovic, A.~Faessler, V.~Rodin, P.~Vogel, and J.~Engel,
 				``Anatomy of nuclear matrix elements for neutrinoless double-beta decay,''
 				Phys.\ Rev.\  C {\bf 77}, 045503 (2008)
 				[arXiv:0710.2055 [nucl-th]].

\bibitem{Wo99}	A.~Wodecki, W.~A.~Kaminski, and F.~\v{S}imkovic,
  				``Grand unified theory constrained supersymmetry and neutrinoless double beta decay,''
  				Phys.\ Rev.\  {\bf D60}, 115007 (1999)
  				[arXiv:hep-ph/9902453].

\bibitem{Gu08}
  				A.~Faessler, T.~Gutsche, S.~Kovalenko, and F.~\v{S}imkovic,
  				``Pion dominance in RPV SUSY induced neutrinoless double beta decay,''
  				Phys.\ Rev.\  D {\bf 77}, 113012 (2008)
  				[arXiv:0710.3199 [hep-ph]].
 
\bibitem{She1}	J.~Menendez, A.~Poves, E.~Caurier and F.~Nowacki,
 				``Disassembling the Nuclear Matrix Elements of the Neutrinoless double beta
 				Decay,''
				  Nucl.\ Phys.\ A {\bf 818}, 139 (2009)
 				[arXiv:0801.3760 [nucl-th]].

\bibitem{Rath} 
 					P.~K.~Rath, R.~Chandra, K.~Chaturvedi, P.~K.~Raina, and J.~G.~Hirsch,
                 ``Deformation effects and neutrinoless positron beta beta decay of Ru-96, 
                 Pd-102, Cd-106, Xe-124, Ba-130 and Dy-156 isotopes within Majorana neutrino 
                 mass mechanism,'' Phys.\ Rev.\  {\bf C80}, 044303 (2009).
                 [arXiv:0906.4476 [nucl-th]].
 
  
\bibitem{De97} D.~S.~Delion, J.~Dukelsky, and P.~Schuck,
				``Restoration of the Ikeda sum rule in self-consistent quasiparticle random-phase approximation,'' 
 				Phys.\ Rev.\ C {\bf 55}, 2340 (1997).
 
\bibitem{Kr98}	F.~Krmpotic, E.~J.~V.~de Passos, D.~S.~Delion, J.~Dukelsky, and P.~Schuck,
  				``Selfconsistent random phase approximation within the O(5) model and Fermi transitions,''
  				Nucl.\ Phys.\ A {\bf 637}, 295 (1998).



\bibitem{Sim08}   F.~\v{S}imkovic, A.~Faessler and P.~Vogel,
 				 ``$0\nu\beta\beta$ nuclear matrix elements and the occupancy of individual orbits,''
				 Phys.\ Rev.\ C {\bf 79}, 015502 (2009)
 				 [arXiv:0812.0348 [nucl-th]].




\bibitem{Ch93}	M.~K.~Cheoun, A.~Bobyk, A.~Faessler, F.~\v{S}imkovic, and G.~Teneva,
  				``Neutron proton pairing in light nuclei and two neutrino double beta decay,''
  				Nucl.\ Phys.\  {\bf A561}, 74-94 (1993).

\bibitem{St09}  F.~\v{S}imkovic, A.~Faessler, H.~Muther, V.~Rodin, and M.~Stauf,
  				``The $0\nu\beta\beta$-decay nuclear matrix elements with self-consistent short-range correlations,''
  				Phys.\ Rev.\  {\bf C79}, 055501 (2009).
  				[arXiv:0902.0331 [nucl-th]].
 
\bibitem{Ba09}	A.~S.~Barabash,
  				``Average and recommended half-life values for two neutrino double beta decay: upgrade-09,''
 				 Proceedings of {\em MEDEX'09}, Workshop on Matrix Elements for the Double-beta-decay EXperiments
  				(Prague, Czech Republic, 2009), ed.\ by O.~Civitarese, I.~Stekl, and J.\ Suhonen,
  				AIP Conf.\ Proc.\ Vol.~{\bf 1180}, 6 (2009).

\bibitem{Over}	A.~Faessler, G.~L.~Fogli, E.~Lisi, V.~Rodin, A.~M.~Rotunno and F.~\v{S}imkovic,
 				``Overconstrained estimates of neutrinoless double beta decay within the QRPA,''
 				J.\ Phys.\ G {\bf 35}, 075104 (2008)
 				[arXiv:0711.3996 [nucl-th]].


\bibitem{Alg1}  J.~W.~Demmel, {\em Applied numerical linear algebra\/} (SIAM, Philadelphia, PA, 1997), 419~pp.

\bibitem{Alg2}	G.~H.~Golub and C.~F.~Van Loan, {\em Matrix computation\/}
					(Johns Hopkins Univ.\ Press, Baltimore, MD, 1996), 728~pp.
					
\bibitem{LAPA}	E.~Anderson {\em et al.}, {\em LAPACK User's Guide\/} (SIAM, Philadelphia, PA, 1994), 407~pp.;
					website: {\tt www.netlib.org/lapack}



\bibitem{Kl04}  H.~V.~Klapdor-Kleingrothaus, I.~V.~Krivosheina, A.~Dietz and O.~Chkvorets,
 				``Search for neutrinoless double beta decay with enriched Ge-76 in Gran
 				Sasso 1990-2003,''
 				Phys.\ Lett.\  B {\bf 586}, 198 (2004)
 				[arXiv:hep-ph/0404088].
				

\bibitem{Kl06}  H.~V.~Klapdor-Kleingrothaus and I.~V.~Krivosheina,
 				``The Evidence For The Observation Of $0\nu\beta\beta$ Decay: The Identification
 				Of $0\nu\beta\beta$ Events From The Full Spectra,''
 				Mod.\ Phys.\ Lett.\  A {\bf 21}, 1547 (2006).

 \bibitem{Bori}   A.~Ali, A.~V.~Borisov, and D.~V.~Zhuridov,
  				``Mechanisms of neutrinoless double-beta decay: A comparative analysis of
  				several nuclei,''
  				Phys.\ Atom.\ Nucl.\  {\bf 73}, 2083 (2010)
  				[Yad.\ Fiz.\  {\bf 73}, 2139 (2010)].

\bibitem{LSSM}   J.~Menendez, A.~Poves, E.~Caurier and F.~Nowacki,
  					``Disassembling the Nuclear Matrix Elements of the Neutrinoless $\beta\beta$ Decay,''
  				Nucl.\ Phys.\  A {\bf 818}, 139 (2009)
  					[arXiv:0801.3760 [nucl-th]].


\bibitem{Iach}   J.~Barea and F.~Iachello,
  			``Neutrinoless double-beta decay in the microscopic interacting boson model,''
  			Phys.\ Rev.\  C {\bf 79}, 044301 (2009). 

\bibitem{HFB1}   
				P.~K.~Rath, R.~Chandra, K.~Chaturvedi, P.~K.~Raina and J.~G.~Hirsch,
  			``Uncertainties in nuclear transition matrix elements for neutrinoless $\beta
  			\beta$ decay within the PHFB model,''
  			Phys.\ Rev.\  C {\bf 82}, 064310 (2010)
  			[arXiv:1104.3965 [nucl-th]].

\bibitem{HFB2}    T.~R.~Rodriguez and G.~Martinez-Pinedo,
  			``Energy density functional study of nuclear matrix elements for neutrinoless
  			$\beta\beta$ decay,''
  			Phys.\ Rev.\ Lett.\  {\bf 105}, 252503 (2010)
  			[arXiv:1008.5260 [nucl-th]].


\bibitem{Fa11}  Amand~Faessler,
                ``Double Beta Decay, Nuclear Structure and Physics beyond the Standard Model,''
                in the Proceedings of {\em NPA5}, International Conference on ``Nuclear Physics in Astrophysics''
                (Eilath, Israel, April 2011), to appear in Journal of Physics (Conference Series);
                arXiv:1104.3700 [nucl-th].

\bibitem{Fo08} 	G.~L.~Fogli, E.~Lisi, A.~Marrone, A.~Melchiorri,  A.~Palazzo, A.~M.~Rotunno, P.~Serra, J.~Silk,
				 and A.~Slosar,
 				``Observables sensitive to absolute neutrino masses. II,''
 				Phys.\ Rev.\  D {\bf 78}, 033010 (2008)
 				[arXiv:0805.2517 [hep-ph]].



\bibitem{Pe11} 	A.~Faessler, A.~Meroni, S.~T.~Petcov,  F.~\v{S}imkovic, and J.~Vergados,
				 ``Uncovering multiple CP-nonconserving mechanisms of $(\beta\beta)_{0\nu}$-Decay,'' 
				 arXiv:1103.2434 [hep-ph].

\bibitem{Supp}    A.~Halprin, S.~T.~Petcov and S.~P.~Rosen,
  				``Effects Of Light And Heavy Majorana Neutrinos In Neutrinoless Double Beta
                 Decay,''
                 Phys.\ Lett.\  B {\bf 125}, 335 (1983).


\end{thebibliography}
\end{document}